\documentclass[conference,letterpaper]{IEEEtran}

\usepackage[utf8]{inputenc}
\usepackage[T1]{fontenc}
\usepackage{url}
\usepackage{ifthen}
\usepackage{cite}
\usepackage[cmex10]{amsmath} % Use the [cmex10] option to ensure complicance
\usepackage{amssymb}
\usepackage{amsthm}	% support \begin{proof}, may consider \begin{IEEEproof} in IEEEtran.cls
\usepackage{color}

\usepackage{kbordermatrix}

\newtheorem{definitionenv}{Definition}
\newtheorem{lemmaenv}[definitionenv]{Lemma}
\newtheorem{theoremenv}[definitionenv]{Theorem}
\newtheorem{corollaryenv}[definitionenv]{Corollary}
\newtheorem{propositionenv}[definitionenv]{Proposition}
\newtheorem{conjectureenv}[definitionenv]{Conjecture}

\newtheorem{app-lemmaenv}[section]{Lemma}
\newtheorem{remarkenv}[definitionenv]{Remark}

\newenvironment{definition}{\begin{definitionenv}\rm}{\end{definitionenv}}
\newenvironment{lemma}{\begin{lemmaenv}\rm}{\end{lemmaenv}}
\newenvironment{theorem}{\begin{theoremenv}\rm}{\end{theoremenv}}
\newenvironment{corollary}{\begin{corollaryenv}\rm}{\end{corollaryenv}}
\newenvironment{proposition}{\begin{propositionenv}\rm}{\end{propositionenv}}
\newenvironment{conjecture}{\begin{conjectureenv}\rm}{\end{conjectureenv}}
\newenvironment{app-lemma}{\begin{app-lemmaenv}\rm}{\end{app-lemmaenv}}
\newenvironment{remark}{\begin{remarkenv}\rm}{\end{remarkenv}}

\newcommand{\bd}{\begin{definition}}
\newcommand{\ed}{\end{definition}}
\newcommand{\edp}{\hspace*{\fill} $\Box$ \end{definition}}
\newcommand{\bl}{\begin{lemma}}
\newcommand{\el}{\end{lemma}}
\newcommand{\elp}{\hspace*{\fill} $\Box$ \end{lemma}}
\newcommand{\bt}{\begin{theorem}}
\newcommand{\et}{\end{theorem}}
\newcommand{\etp}{\hspace*{\fill} $\Box$ \end{theorem}}
\newcommand{\bc}{\begin{corollary}}
\newcommand{\ec}{\end{corollary}}
\newcommand{\ecp}{\hspace*{\fill} $\Box$ \end{corollary}}
\newcommand{\bcj}{\begin{conjecture}}
\newcommand{\ecj}{\end{conjecture}}
\newcommand{\be}{\begin{example}}
\newcommand{\ee}{\end{example}}
\newcommand{\eep}{\hspace*{\fill} $\Box$ \end{example}}
\newcommand{\bp}{\begin{proposition}}
\newcommand{\ep}{\end{proposition}}
\newcommand{\epp}{\hspace*{\fill} $\Box$ \end{proposition}}
\newcommand{\br}{\begin{remark}}
\newcommand{\er}{\end{remark}}
\newcommand{\erp}{\hspace*{\fill} $\Box$ \end{remark}}

\DeclareMathOperator{\Row}{Row}

\newcommand{\fa} {~\forall~}
\newcommand{\eq}[1]{\eqref{#1}}       % IEEE

\newcommand{\ket}[1]{|#1\rangle}

\newcommand{\beq}[1]{\begin{equation*}{ #1 }\end{equation*}}
\newcommand{\beql}[2]{\begin{equation}\label{#1}{ #2 }\end{equation}}
\newcommand{\beqs}[1]{\begin{align*} #1 \end{align*}}
\newcommand{\beqsl}[2]{\begin{equation}\label{#1}\begin{aligned} #2 \end{aligned}\end{equation}}

\newcommand{\bcase}[1]{\begin{cases} #1 \end{cases}}
\newcommand{\bmat}[2]{\left[ \begin{array}{#1}  #2  \end{array} \right]}
\newcommand{\bmtx}[1]{\left[ \begin{matrix}  #1  \end{matrix} \right]}
\newcommand{\bsmtx}[1]{\left[ \begin{smallmatrix} #1 \end{smallmatrix} \right]}
\newcommand{\ob}[1]{\overbrace{}^{#1}}

\newcommand{\CC}{{\mathbb C}}
\newcommand{\ZZ}{{\mathbb Z}}

\newcommand{\hE}{{\hat E}}

\newcommand{\bX}{{\bar X}}
\newcommand{\bZ}{{\bar Z}}
\newcommand{\sC}{{\cal C}}

\newcommand{\sG}{{\cal G}}

\newcommand{\sS}{{\cal S}}

\newcommand{\wt}{\mathrm{wt}}
\newcommand{\gw}{\mathrm{gw}}

\newcommand{\af}{\alpha}

\newcommand{\sig}{\sigma}

\newcommand{\La}{\Lambda}
\newcommand{\dg}{\dagger}

\newcommand{\teq}{\triangleq}
\newcommand{\ceil}[1]{\left\lceil {#1} \right\rceil}
\newcommand{\floor}[1]{\left\lfloor {#1} \right\rfloor}

\makeatletter
\newif\if@borderstar
\def\bordermatrix{\@ifnextchar*{%
  \@borderstartrue\@bordermatrix@i}{\@borderstarfalse\@bordermatrix@i*}}
\def\@bordermatrix@i*{\@ifnextchar[{\@bordermatrix@ii}{\@bordermatrix@ii[()]}}
\def\@bordermatrix@ii[#1]#2{%
  \begingroup
  \m@th\@tempdima8.75\p@\setbox\z@\vbox{%
    \def\cr{\crcr\noalign{\kern 2\p@\global\let\cr\endline }}%
    \ialign {$##$\hfil\kern 2\p@\kern\@tempdima & \thinspace %
      \hfil $##$\hfil && \quad\hfil $##$\hfil\crcr\omit\strut %
      \hfil\crcr\noalign{\kern -\baselineskip}#2\crcr\omit %
      \strut\cr}}%
  \setbox\tw@\vbox{\unvcopy\z@\global\setbox\@ne\lastbox}%
  \setbox\tw@\hbox{\unhbox\@ne\unskip\global\setbox\@ne\lastbox}%
  \setbox\tw@\hbox{%
    $\kern\wd\@ne\kern -\@tempdima\left\@firstoftwo#1%
      \if@borderstar\kern2pt\else\kern -\wd\@ne\fi%
    \global\setbox\@ne\vbox{\box\@ne\if@borderstar\else\kern 2\p@\fi}%
    \vcenter{\if@borderstar\else\kern -\ht\@ne\fi%
      \unvbox\z@\kern-\if@borderstar2\fi\baselineskip}%
    \if@borderstar\kern-2\@tempdima\kern2\p@\else\,\fi\right\@secondoftwo#1 $%
  }\null \;\vbox{\kern\ht\@ne\box\tw@}%
  \endgroup
}

\begin{document}
\title{The Encoding and Decoding Complexities of
	Entanglement-Assisted Quantum Stabilizer Codes}

%%% Several authors with up to three affiliations:
\author{%
  \IEEEauthorblockN{Kao-Yueh Kuo}
  \IEEEauthorblockA{Institute of Communications Engineering\\
			National Chiao Tung University, Hsinchu 30010, Taiwan\\
                    Email:  kykuo@nctu.edu.tw}
  \and
  \IEEEauthorblockN{Ching-Yi Lai}
  \IEEEauthorblockA{Institute of Communications Engineering\\
			National Chiao Tung University, Hsinchu 30010, Taiwan\\
                    Email: cylai@nctu.edu.tw}
}
%%% Many authors with many affiliations:
% \author{%
%   \IEEEauthorblockN{Albus Dumbledore\IEEEauthorrefmark{1},
%                     Olympe Maxime\IEEEauthorrefmark{2},
%                     Stefan M.~Moser\IEEEauthorrefmark{3}\IEEEauthorrefmark{4},
%                     and Harry Potter\IEEEauthorrefmark{1}}
%   \IEEEauthorblockA{\IEEEauthorrefmark{1}%
%                     Hogwarts School of Witchcraft and Wizardry,
%                     1714 Hogsmeade, Scotland,
%                     \{dumbledore, potter\}@hogwarts.edu}
%   \IEEEauthorblockA{\IEEEauthorrefmark{2}%
%                     Beauxbatons Academy of Magic,
%                     1290 Pyrénées, France,
%                     maxime@beauxbatons.edu}
%   \IEEEauthorblockA{\IEEEauthorrefmark{3}%
%                     ETH Zürich, ISI (D-ITET), ETH Zentrum,
%                     CH-8092 Zürich, Switzerland,
%                     moser@isi.ee.ethz.ch}
%   \IEEEauthorblockA{\IEEEauthorrefmark{4}%
%                     National Chiao Tung University (NCTU),
%                     Hsinchu, Taiwan,
%                     moser@isi.ee.ethz.ch}
% }

\maketitle

%%%%%%
%% Abstract:
%% If your paper is eligible for the student paper award, please add
%% the comment "THIS PAPER IS ELIGIBLE FOR THE STUDENT PAPER AWARD."
%% as a first line in the abstract. For the final version of the accepted paper, %% please do not forget to remove this comment!
%%
\begin{abstract}
%Lower encoding complexity implies a faster overall fault-tolerant computation to maintain the decay-with-time quantum coherence.
Quantum error-correcting codes are used to protect quantum information from decoherence. A raw state is mapped,  by an encoding circuit, to a codeword so that the most likely quantum errors from a noisy quantum channel can be removed after a decoding process.

A good encoding circuit should have some desired features, such as low depth, few gates, and so on. In this paper, we show how to practically implement an encoding circuit of gate complexity $O(n(n-k+c)/\log n)$ for an $[[n,k;c]]$ quantum stabilizer code with the help of $c$ pairs of maximally-entangled states. 
For the special case of an $[[n,k]]$ stabilizer code with $c=0$, the encoding complexity is $O(n(n-k)/\log n)$, which is previously known to be $O(n^2/\log n)$.
For $c>0,$ this suggests that the benefits from shared entanglement come at an additional cost of  encoding complexity.

{
Finally we discuss decoding of entanglement-assisted quantum stabilizer codes and extend   previously known computational hardness results on decoding quantum stabilizer codes.} 
\end{abstract}

%% The paper must be self-contained. However, if you are referring to
%% a full version for checking certain proofs, please provide the
%% publically accessible location below.  If the paper is completely
%% self-contained, you can remove the following line from your submission.
%\textit{A full version of this paper is accessible at:}
%\url{http://isit2019.fr/}

\section{Introduction} \label{sec:Intro}
%%% IEEE journal starts like
%\IEEEPARstart{Q}{uantum} information ...
Quantum computers are powerful. However they are hard to build because  quantum states are vulnerable and physical  gates are imperfect. How to handle quantum noises has been an important problem in quantum information processing.  A possible method is to use quantum error-correcting   codes \cite{Shor95,CS96,Steane96}, in which quantum information is encoded in a codespace so that the most likely errors can be treated.

  The class of quantum stabilizer codes (QSCs) have similar features to  classical linear codes and are convenient for practical implementations~\cite{GotPhD}. An $[[n,k]]$ QSC is a $k$-qubit subspace of the $n$-qubit state space. The mapping from the raw $k$-qubit space to the encoded space can be implemented by an encoding circuit consisting of elementary gates. Clearly a low-complexity encoding circuit is desired, since quantum coherence decays with time and quantum gates cannot be implemented perfectly. In addition, the syndrome measurement circuit for error correction is also closely related to the encoding circuit \cite{NC00}. As a consequence, a low complexity encoding circuit can be potentially used in designing fault-tolerant procedures to achieve higher error threshold. %in fault-tolerant quantum computation.

In this paper we study the gate complexity of an encoding circuit. %is counted by the number of used elementary gates.
To encode an $[[n,k]]$ QSC, Cleve and Gottesman showed that $O(n(n-k))$ gates are required \cite{CG96},
which is proportional to the dimension  of the underlying \emph{check matrix}. % of the underlying stabilizer code's check matrix.
Aaronson and Gottesman argued that  $O(n^2/\log n)$ gates are sufficient~\cite{AG04}
by using the CNOT circuit decomposition algorithm in~\cite{PMH08}.
Herein we  show that this  complexity can be further reduced to $O(n(n-k)/\log n)$ %for an $[[n,k]]$ stabilizer code,
by studying the structure of the check matrix and applying a variant of the decomposition algorithm~in~\cite{PMH08}.
More generally, we show that the encoding complexity for an $[[n,k;c]]$ entanglement-assisted  quantum stabilizer code~(EAQSC)~\cite{BDH06}
 is  $O(n(n-k+c)/\log n)$. 
 It is known that entanglement may increase the code rate~\cite{BDH06}, provide a larger minimum distance \cite{LB13}, or improve the decoding performance~\cite{HBD09}. It is clear now that these benefits come at an additional cost of encoding compleity. 
  %but it will also need larger encoding complexity $O(n(n-k+c)/\log n)$ as we will show here. 

%On the other hand, it desires to have a very hard general decoding problem to construct a code-based \emph{McEliece cryptosystem} \cite{McE78}. 
On the other hand, decoding a linear code is shown to be NP-complete by Berlekamp, McEliece, and van Tilborg \cite{BMT78}. 
 In the quantum case, several decoding problems are also shown to be NP-hard \cite{HG10,Fuj12,KL13}.
 Especially in the presence of degeneracy, the optimal decoding method finds the most probable equivalent coset of errors corresponding to an error syndrome and it is shown to be \#P-complete by Iyer and Poulin~\cite{IP15}.
  In this paper we will discuss the decoding procedure of a general EAQSC and show that the corresponding decoding problem is also \#P-complete.

This paper is organized as follows. In Section~\ref{sec:StdForm} we give the notation and the basics of quantum codes.
The encoding complexity of $O(n(n-k+c)/\log n))$ is shown in Section~\ref{sec:EncComp}. 
 {In Section~\ref{sec:DecComp}, we discuss the hardness of decoding  an EAQSC.} Then we conclude.

\section{The EAQSC Scheme} \label{sec:StdForm}
We first review basics of stabilizer codes. 
Let $\sG_n$ be the $n$-fold Pauli group $\{ i^e \bigotimes_{j=1}^n M_j: M_j\in\{I,X,Y,Z\}, e\in\{0,1,2,3\}\}$.
Let $\sS\subset \sG_n$ be an \emph{Abelian} stabilizer group, generated by $n-k$ independent \emph{stabilizer generators}
$\bZ_1, \dots, \bZ_{n-k}$, such that $-I \notin\sS$. 
Then an $[[n,k]]$ stabilizer code defined by $\sS$ is $\sC(\sS) \teq \{\ket{v} \in \CC^{2^n}:~ {g\ket{v} = \ket{v}} ~~\forall~ g \in \sS\}$.

Let $Z_i \teq I^{\otimes(i-1)} \otimes Z \otimes I^{\otimes(n-i)}$ and similarly for $X_i$. 
For ${x}= (x_1 \cdots x_k)\in\{0,1\}^n$, consider the following initial state
$$\ket{0}^{\otimes n-k} \ket{x}= X_{n-k+1}^{x_1} \cdots X_{n}^{x_k} \ket{0}^{\otimes n}$$
with $k$-qubit logical state $\ket{x}$ and $n-k$ ancillas in $\ket{0}$ before encoding. 
This state has stabilizers $Z_1,\dots, Z_{n-k}$.
A unitary encoding circuit $U$ for $\sC(\sS)$ maps $\ket{x}$ to $\ket{\bar{x}} \teq U \ket{0}^{\otimes n-k}\ket{x}$ 
with stabilizers $\bZ_i = U Z_i U^\dg,~i=1,\dots,n-k$.
The set of the $2^k$ vectors $\{ \ket{\bar{x}}: x\in\{0,1\}^n \}$ forms a basis of $\sC(\sS)$.
Also $\ket{\bar{x}}$ can be written as $\bX_{n-k+1}^{x_1} \cdots \bX_{n}^{x_k} \ket{\bar{0}},$ where 
$\bX_{j} = U X_j U^\dg,~j=n-k+1,\dots,n,$ are called \emph{seed generators} 
and $\ket{\bar{0}} = U\ket{0}^{\otimes n}$ is a \emph{coded zero}, which can be generated by
$\ket{\bar{0}} = \frac{1}{\sqrt{2^{n-k}}} \prod_{i=1}^{n-k}(I + \bZ_i) \ket{0}^{\otimes n} 
			= \frac{1}{\sqrt{2^{n-k}}} \sum_{g\in \sS} g\ket{0}^{\otimes n}$
such that $g\ket{\bar{0}} = \ket{\bar{0}}$ for all $g\in \sS$. 

Define a homomorphism $\varphi: \sG_n\to \{0,1\}^{2n}$ by $$\varphi(g)\triangleq (u_1\cdots u_n|v_1\cdots v_n)$$ 
for $g = \sig_1 \otimes\cdots\otimes\sig_n\in \sG_n$ such that $\sig_j\mapsto(u_j|v_j)$ 
by $I\mapsto(0|0),X\mapsto(1|0),Y\mapsto(1|1),Z\mapsto(0|1)$.
An $(n-k)\times 2n$ binary check matrix $H = [I~A|B~C]$ is the matrix with rows $\varphi(\bZ_i),~i=1,\dots, n-k$.
For example, the initial state has a check matrix $[O~O|I~O]$, where $O$ is the all-zero matrix of appropriate dimensions.
Since  $\sS$ is an Abelian subgroup, $H$ has to satisfy
the commutation condition $H\La H^T = O$, where
$\La = \left[
  {O_{n \times n} \atop I_{n \times n}} \vrule ~
  {I_{n \times n} \atop O_{n \times n}} \right].
$
 The encoding circuit can be implemented  by the reverse of a series of elementary gates that 
transform $H$ to $[O~O|I~O]$~\cite{CG96,NC00,AG04}. %, which represents $\varphi(Z_i), i=1,\cdots, n-k$,

EAQSCs
are a coding scheme that  the sender (Alice) and the receiver (Bob) share some maximally-entangled Einstein-Podolsky-Rosen (EPR) pairs $\ket{\Phi^+}_{AB}$  \cite{BDH06,LA18}.
%Let $\ket{\Phi^+}^{\otimes c}_{AB}$ be the state of $c$ EPR pairs ($c$ ebits, consisting of $2c$ qubits) pre-shared between Alice and Bob. %Let $s=n-k-c$.
An initial basis state of the overall system is
$\ket{0}^{\otimes s} \otimes \ket{\Phi^+}^{\otimes c}_{AB}\otimes \ket{x_1\cdots x_k}$,
where Bob holds $c$ qubits (a half the $c$ EPR pairs) and Alice holds the remaining $n=s+c+k$ qubits prior to communication.
After encoding, Alice sends her $n$ qubits to Bob through a noisy quantum channel and Bob's $c$ qubits are assumed to be error-free.
Every initial basis state is stabilized by a set of operators in $\sG_{n+c}$ with corresponding check matrix (before encoding) 
%%keep a next empty row

$H_{\text{raw}} =$
\small
\beq{
  \kbordermatrix{
  \mbox{~}&\ob{s}&\ob{c}&\ob{c}&\ob{k} & &\ob{s}&\ob{c}&\ob{c}&\ob{k}\\
   &O &O &O             &O             &\vrule &I_{s\times s} &O &O             &O  \\
   &O &I_{c\times c} &I_{c\times c} &O &\vrule &O             &O &O             &O  \\
   &O &O &O             &O             &\vrule &O             &I_{c\times c}&I_{c\times c} &O
  }.
}
\normalsize
Similarly, we can consider a unitary encoding circuit $U$ that maps $H_{\text{raw}}$ to some $H$ with $H\La H^T=O$. Upon reception, Bob  does a decoding on the total of $(n+c)$ qubits according to the check matrix $H$. 
%Here Bob performs a regular stabilizer decoding, and the errors in the first $n$ qubits caused by communication could be corrected by the parity-check conditions defined in $H$.

We can consider a \emph{simplified check matrix} of $H_{\text{raw}}$ without the columns corresponding to Bob's qubits:
\small
\beql{H'_raw}{
H'_{\text{raw}} =
  \kbordermatrix{
  \mbox{~}&\ob{s}&\ob{c}&\ob{k} & &\ob{s}&\ob{c}&\ob{k}\\
    s\big\{ &O &O             &O &\vrule &I_{s\times s} &O             &O \\
    c\big\{ &O &I_{c\times c} &O &\vrule &O             &O             &O \\
    c\big\{ &O &O             &O &\vrule &O             &I_{c\times c} &O
  },
}
\normalsize
which is corresponding to \emph{simplified stabilizers} 
$\{Z_1, \dots, Z_s,$ $X_{s+1}, \dots, X_{s+c},$ $Z_{s+1}, \dots, Z_{s+c}\}$ in $\sG_n$.
After encoding by $U$, we have simplified stabilizer generators $\bZ_i = UZ_i U^\dg$ and $\bX_i = UX_i U^\dg$ satisfying the following commutation relations
\beqsl{ccs}{ ~ % don't know why without this space ~ the part of 1st row vanish
[\bZ_i, \bZ_j] &= 0, \quad \fa i,j; \\
[\bX_i, \bX_j] &= 0, \quad \fa i,j; \\
[\bZ_i, \bX_j] &= 0, \quad \fa i\ne j; \\
\{\bZ_i, \bX_i\} &= 1, \quad \fa i = s+1, \dots, s+c.
}
Thus we have a \emph{non-Abelian} subgroup
$\sS' = \langle\bZ_1, \dots, \bZ_s,$ $\bX_{s+1}, \dots, \bX_{s+c},$ $\bZ_{s+1}, \dots, \bZ_{s+c}\rangle$
%with the generators, if using the binary representation, having a \emph{standard form} for encoding
with corresponding simplified check matrix
\beqsl{H'}{
&H'_{(s+2c) \times 2n} = \\
&~~\kbordermatrix{
  \mbox{~}&\ob{s}&\ob{c+k} & &\ob{s}&\ob{c+k}\\
    s\big\{ & I & A    &\vrule & B   & C   \\
    c\big\{ & O & M_1  &\vrule & M_5 & M_2 \\
    c\big\{ & O & M_3  &\vrule & M_6 & M_4
  }
=
\left[ \begin{smallmatrix}
  \varphi(\bZ_1) \\[-6pt]
  \vdots \\
  \varphi(\bZ_s) \\
  \varphi(\bX_{s+1}) \\[-6pt]
  \vdots \\
  \varphi(\bX_{s+c}) \\
  \varphi(\bZ_{s+1}) \\[-6pt]
  \vdots \\
  \varphi(\bZ_{s+c}) \\
  \end{smallmatrix} \right].
}
By \eq{ccs} we have
\beqsl{ccs0}{
  & B+CA^T + B^T+AC^T = O,		\\ %\label{ccB}\\
  & M_5 = M_1 C^T + M_2 A^T,	\\ %\label{cc5}\\
  & M_6 = M_3 C^T + M_4 A^T,	\\ %\label{cc6}\\
  & M_1 M_2^T + M_2 M_1^T = O,	\\ %\label{ccZ}\\
  & M_3 M_4^T + M_4 M_3^T = O,	\\ %\label{ccX}\\
  & M_1 M_4^T + M_2 M_3^T = I.	   %\label{ac}
}
Without loss of generality, we assume $M_1=[M_{11}~M_{12}]$ with non-singular $M_{11}$.
Now a \emph{standard form} of the check matrix (with additional columns corresponding Bob's qubits) is 
\beqsl{hH}{
&H_{(s+2c) \times 2(n+c)} = \\
&\quad \kbordermatrix{
  \mbox{~}&\ob{s}&\ob{c+k}&\ob{c} & &\ob{s}&\ob{c+k}&\ob{c}\\
    s\big\{ & I & A  & O  &\vrule & B   & C   & O  \\
    c\big\{ & O & M_1& I  &\vrule & M_5 & M_2 & O  \\
    c\big\{ & O & M_3& O  &\vrule & M_6 & M_4 & I
  },
}
which satisfies $H\La H^T = O$.

\section{Encoding complexity} \label{sec:EncComp}
We consider the decomposition of a stabilizer circuit~\cite{AG04} by Clifford gates \{CNOT, H, P\} and (possibly) some swap operations.
(See, for example,~\cite{Wilde10} for the encoding cirucit of an EAQSC.)
  %Note that in the following we use CNOT, H, P,
  For convenience, we use an additional controlled-$Z$ gate (CZ), which can be decomposed as a CNOT and two H gates, in the circuit decomposition. 
Consider a check matrix of the form $H= [H_X|H_Z]$. We have the following gate operation rules:
\begin{enumerate}
\item A CNOT gate from qubit $i$ to qubit $j$
	adds column $i$ to column $j$ in $H_X$ and
	adds column $j$ to column $i$ in $H_Z$.
\item A  CZ gate from qubit $i$ to qubit $j$
	adds column $i$ in $H_X$ to column~$j$ in $H_Z$ and
	adds column $j$ in $H_X$ to column~$i$ in $H_Z$.
\item A Hadamard gate on qubit~$i$
	swaps column $i$ in $H_X$ with column~$i$ in $H_Z$.
\item A phase gate on qubit~$i$
	adds column $i$ in $H_X$ to column~$i$ in $H_Z$.
\end{enumerate}
Patel, Markov, and Hayes proposed an efficient reduction algorithm to decompose a CNOT circuit~\cite{PMH08}.
Motivated by their method, we propose  the following algorithm for the reduction of a matrix of the form $[I\ A | B' \ O]$ for our purpose.
\bl \label{IA} %%%%%%%%%%%%%%%%%%%%%%%%%%%%%%
For an $(n-k)\times 2n$ check matrix $H=[I\ A | B'\ O]$, there exists a linear transformation that
maps $H$ to $[I\ O| B'\ O]$ using $O\left((2^m+k) \cdot \frac{n-k}{m}\right)$ CNOT gates for some $m$ smaller than $n-k$.
\el %%%%%%%%%%%%%%%%%%%%%%%%%%%%%%%%%%%
\begin{proof}
For simplicity, we first assume that $n-k$ is divisible by some integer $m$.
Partition $[I~A]$ into $(n-k)/m$ blocks
\small
\beqs{
\begin{array}{c}
  \text{1st block}\cr
  \vdots\cr
  \frac{n-k}{m}\text{th block}\cr
\end{array}
\
\bmtx{
  I_{m\times m} &       &                &A_1    \cr
                &\ddots &                &\vdots \cr
                &       &I_{m\times m}   &~A_{\frac{n-k}{m}} \cr
  },
}
\normalsize
where   $A_i$ is an $m\times k$ binary matrix.
Beginning with $i=1$, %the first section,
we perform appropriate column operations so that
the columns of $I_{m\times m}$ in the $i$th block 
 will go over all nonzero $m$-bit vectors and transform to a matrix $D_{m\times m}$
with all ones in the upper triangular part. For example, if $m=3$,
%\small
\beqsl{eqn:UpTriMtx}{
I_{m\times m} &=
\left[ \begin{smallmatrix}
  1 & 0 & 0 \cr
  0 & 1 & 0 \cr
  0 & 0 & 1 \cr
\end{smallmatrix} \right]
\to
\left[ \begin{smallmatrix}
  1 & 1 & 0 \cr
  0 & 1 & 0 \cr
  0 & 0 & 1 \cr
\end{smallmatrix} \right]
\to
\left[ \begin{smallmatrix}
  1 & 1 & 1 \cr
  0 & 1 & 0 \cr
  0 & 0 & 1 \cr
\end{smallmatrix} \right]
\to
\left[ \begin{smallmatrix}
  1 & 1 & 0 \cr
  0 & 1 & 1 \cr
  0 & 0 & 1 \cr
\end{smallmatrix} \right] \\
&\to
\left[ \begin{smallmatrix}
  1 & 1 & 1 \cr
  0 & 1 & 1 \cr
  0 & 0 & 1 \cr
\end{smallmatrix} \right]
=D_{m\times m}.
}
\normalsize
This requires $2^m-m-1$ operations per block.
%, and all non-zero column vectors of length $m$ are generated during this process.
%
During this process,
if there is any column in $A_i$ that is identical to a generated column,
we can eliminate that column and this takes at most $k$ operations (since $A_i$ has $k$ columns).
Next, we transform $D_{m\times m}$ back to $I_{m\times m}$,
and this needs $m-1$ operations.
For example, consider $A_i = \bsmtx{1 &1 &1 &1 \cr 1 &1 &1 &1 \cr 1 &0 &0 &1 \cr}$:
\beqs{
[I\ A_i] =
&\left[ \begin{smallmatrix}
  1 & 0 & 0 &&1 &1 &1 &1 \cr
  0 & 1 & 0 &&1 &1 &1 &1 \cr
  0 & 0 & 1 &&1 &0 &0 &1 \cr
\end{smallmatrix} \right]
\to
\left[ \begin{smallmatrix}
  1 & 1 & 0 &&1 &1 &1 &1 \cr
  0 & 1 & 0 &&1 &1 &1 &1 \cr
  0 & 0 & 1 &&1 &0 &0 &1 \cr
\end{smallmatrix} \right]
\to
\left[ \begin{smallmatrix}
  1 & 1 & 0 &&1 &0 &0 &1 \cr
  0 & 1 & 0 &&1 &0 &0 &1 \cr
  0 & 0 & 1 &&1 &0 &0 &1 \cr
\end{smallmatrix} \right] \\
\to
&\left[ \begin{smallmatrix}
  1 & 1 & 1 &&1 &0 &0 &1 \cr
  0 & 1 & 1 &&1 &0 &0 &1 \cr
  0 & 0 & 1 &&1 &0 &0 &1 \cr
\end{smallmatrix} \right]
\to
\left[ \begin{smallmatrix}
  1 & 1 & 1 &&0 &0 &0 &0 \cr
  0 & 1 & 1 &&0 &0 &0 &0 \cr
  0 & 0 & 1 &&0 &0 &0 &0 \cr
\end{smallmatrix} \right]
\to
\left[ \begin{smallmatrix}
  1 & 0 & 0 &&0 &0 &0 &0 \cr
  0 & 1 & 0 &&0 &0 &0 &0 \cr
  0 & 0 & 1 &&0 &0 &0 &0 \cr
\end{smallmatrix} \right].
}
So it takes at most $((2^m-m-1) + k + (m-1)) < (2^m+k)$ gates to eliminate one block.
By repeating the above process for the $\ceil{\frac{n-k}{m}}$ blocks, it requires at most
$(2^m+k) \ceil{\frac{n-k}{m}} = O((2^m+k)\cdot\frac{n-k}{m})$
gates to eliminate $A$. %because $\ceil{\frac{n-k}{m}} \le 2(\frac{n-k}{m})$.
\end{proof}
Similarly we propose the following decomposition algorithm for certain phase and CZ circuits.
\bl \label{IB} %%%%%%%%%%%%%%%%%%%%%%%%%
For an $(n-k)\times 2n$ check matrix $H=[I\ O | B'\ O]$ with   symmetric $B'$, there exists a linear transformation
that maps $H$ to $[I\ O| O\ O]$ using  {$O\left((2^m + n-k) \cdot \frac{n-k}{m}\right)$} gates.
\el %%%%%%%%%%%%%%%%%%%%%%%%%%%%%%%%%%%%%
\begin{proof}
By Rule 4), it needs at most $n-k = O(n-k)$ phase gates to eliminate non-zero diagonal entries in $B'$.
After that, CZ gates can further reduce $H$ to $[I\ O| O\ O]$ by Rule 2).
To decompose an efficient CZ circuit, after phase gates, we first reduce $B'$ to
\small
\beql{B_0}{
B_0=
\left[ \begin{matrix}
  O_{m\times m} & *              & \cdots & * \cr
  *             & O_{m\times m}  &*\cdots & * \cr
  \vdots        & *\atop{\vdots} & \ddots & \vdots \atop * \cr
  *             & *              &\cdots * & O_{m\times m} \cr
\end{matrix} \right]
\begin{array}{c}
  \text{1st block}\cr
  \cdot\cr \cdot\cr \cdot\cr
  \frac{n-k}{m}\text{th block}.\cr
\end{array}
}
\normalsize
To obtain $O_{m\times m}$, every block needs fewer than $m^2$ CZ gates.
To eliminate the * parts, every block needs fewer than ${(n-k)}$ CZ gates %(actually fewer than $(n-k)/2$ gates)
by the same technique as in Lemma~\ref{IA}  (though additional $2^m$ CNOT gates are required).
Thus eliminating $B'$ needs a gate complexity
\small
\beqs{
&\quad
\underbrace{O(n-k)}_{\text{phase gates}} + \underbrace{O\left(2^m\cdot\frac{n-k}{m}\right)}_{\text{CNOT gates}}
  + \underbrace{O\left((m^2 + n-k)\cdot\frac{n-k}{m}\right) }_{\text{CZ gates}} \\
&= O\left((m+2^m +m^2+ n-k)\frac{n-k}{m}\right)
~= O\left((2^m + n-k)\frac{n-k}{m}\right).
}
\normalsize
\end{proof} %%%%%%%%%%%%%%%%%%%%%%%%%%%%%%%%%%%%%%%%%%%%%
By the two lemmas and \eq{ccs0}, we have the following theorems. 
\bt \label{IA_IC} %%%%%%%%%%%%%%%%%%%%%%%%%%%%%%%%%%
For an $(n-k)\times 2n$ check matrix $H=[I\ A | B\ C]$ and some $m$ smaller than $n-k$,
there exists a linear transformation that maps $H$ to $[I\ O| B'\ C]$
with $B' = B+CA^T$ using $O\left((2^m+k) \cdot \frac{n-k}{m}\right)$ CNOT gates,
and there exists a linear transformation that maps $H$ to $[I\ A| B''\ O]$
with $B'' = B+AC^T$ using $O\left((2^m+k) \cdot \frac{n-k}{m}\right)$ gates, %CZ gates; % (in general need CNOT gates for the $2^m$)
where  $B'$ and $B''$ are both symmetric.
\et
%\begin{proof} %%%%%%%%%%%%%%%%%%%%%%%%%%%%%%%%%%%%
%By Lemma \ref{IA} and Rule 1) above, the first part of this theorem holds.
%By Lemma \ref{IA} and Rule 2) above, the second part of this theorem holds.
%The third part of this theorem holds by \eq{ccs0}.
%\end{proof} %%%%%%%%%%%%%%%%%%%%%%%%%%%%%%%%%%%%%%%%%%%%%
%
\bt \label{O(H)} %%%%%%%%%%%%%%%%%%%%%
	An $[[n,k]]$ QSC has an encoding complexity $O(n(n-k)/\log n)$.
\et
\begin{proof}
Without loss of generality, let 
$H = [I~A|B~C]_{(n-k)\times 2n}$ be the check matrix of a target stabilizer code.
 We   add $k$ seed generators for the purpose of elimination as in \cite{GotPhD}, \cite{KL10},
and their corresponding binary matrix can be represented by $[O~I | C^T O]_{k\times 2n}$.
The procedure of reduction is as follows.
\small
\beqs{
&\kbordermatrix{
  \mbox{~}&\ob{n-k}&\ob{k} & &\ob{n-k}&\ob{k}\\
    n-k\big\{ & I & A  &\vrule & B   & C  \\
    ~~~k\big\{ & O & I  &\vrule & C^T & O
  }\\[5pt]
%& \bmat{cc|cc}{
%  I & A   & B   & C\\
%  O & I   & C^T & O
%  }\\
& \to
  \bmat{cc|cc}{
  I & A   & B' & O\\
  O & I   & O & O
  }
  ~{\text{by Theorem~\ref{IA_IC}, where $C$ is eliminated}
     \atop
     \text{by $O((2^m+k)\cdot\frac{n-k}{m})$ gates and}
    }\\
  &~~~~~~~~~~~~~~~~~~~~~~~~~~~~~~~~~~~\text{$B' = B+AC^T$ is symmetric;}\\
  &~~~~\text{ $C^T$ is also eliminated without additional gates by Rule 2);} \\[5pt]
& \to
  \bmat{cc|cc}{
  I & O   & B' & O\\
  O & I   & O  & O
  }
  ~{\text{by Lemma~\ref{IA}, where $A$ is eliminated by}
     \atop
     \text{$O((2^m+k)\cdot\frac{n-k}{m})$ gates;~~~~~~~~~~~~}
    }\\[5pt]
& \to
  \bmat{cc|cc}{
  I & O   & O & O\\
  O & I   & O & O
  }
  ~{\text{by Lemma~\ref{IB}, where $B'$ is eliminated by}
     \atop
     \text{$O((2^m+n-k)\cdot\frac{n-k}{m})$ gates;~~~~~~}
    }\\[5pt]
& \to
  \bmat{cc|cc}{
  O & O   & I & O\\
  O & I   & O & O
  }~~\text{by $O(n-k)$ Hadamard gates (Rule 3))}.
}
\normalsize
Note that Hadamard gates on the last $k$ qubits are not required since the last $k$ rows correspond to the seed generators~\cite{CG96}.
%The  reduction can be focused on $H$ without putting on the seed generators. 
The overall  complexity is
\small
\beqs{
&2\cdot O\left((2^m+k) \cdot \frac{n-k}{m}\right) + O\left((2^m+ n-k) \cdot \frac{n-k}{m}\right) + O(n-k) \\
&= O\left((2^m +n +m)\cdot\frac{n-k}{m}\right) = O\left((2^m +n)\cdot\frac{n-k}{m}\right).
}
\normalsize
Similar to \cite{PMH08}, taking $m = \floor{\af\log_2 n}$ for some $\af <1$ and we have
  $O((2^m +n)\cdot\frac{n-k}{m}) = O(n\cdot\frac{n-k}{\log n})$
since ${2^m\leq n^\af =o(n)}$.
\end{proof}
The  lemmas and theorems in this section are based on a trick:  Originally it costs $O(n)$ gates to reduce each of the $(n-k)$ rows, but if we group every $m$ rows as a block, then we only have $O((n-k)/m)$ blocks and every block needs at most $O(2^m + n) = O(n)$ gates.
 %by carefully selecting $m$. %and using the elimination skill in the proof.
%
%Similar inference can be applied to a full row-rank matrix defined by \eq{hH}.
\bt \label{O(hH)} %%%%%%%%%%%%%%%%%%%%%%%%%%%%%%%%%%%%%%%%%%
An $[[n,k;c]]$ EAQSC has an encoding complexity $O(n(n-k+c)/\log n)$.
\et
\begin{proof}
It suffices to show that 
the required number of gates is proportional to $(n-k+c) \times 2n$
to reduce  \eq{H'} to \eq{H'_raw}.
% before sectionization.
By \eq{ccs0} and  Rules 1) to 4) above, we start to reduce \eq{H'} to \eq{H'_raw}:
\small
\beqs{
  &\bmat{cc|cc}{
  I & A     &  B               & C   \\
  O & M_1   &  M_1C^T + M_2A^T & M_2 \\
  O & M_3   &  M_3C^T + M_4A^T & M_4
  }\\%_{(s+2c)\times 2n}%,~\text{assuming $M_1=[M_{11}~M_{12}]$, $M_{11}$ non-singular}\\
%  &\kbordermatrix{
%    \mbox{~}&\ob{s}&\ob{c+k} & &\ob{s}&\ob{c+k}\\
%      s\big\{ & I & A    &\vrule & B   & C     \\
%      c\big\{ & O & M_1  &\vrule & M_5 & M_2   \\
%      c\big\{ & O & M_3  &\vrule & M_6 & M_4
%    }
%  \\
  \to &
  \bmat{cc|cc}{
  I & O     &  B+CA^T & C   \\
  O & M_1   &  M_1C^T & M_2 \\
  O & M_3   &  M_3C^T & M_4
  }
  ~~\text{by Theorem~\ref{IA_IC};}\\
  \to &
  \bmat{cc|cc}{
  I & O     &  B+CA^T & O   \\
  O & M_1   &  O      & M_2 \\
  O & M_3   &  O      & M_4
  }
  ~~\text{ by Theorem~\ref{IA_IC} again;}\\
%  \\
%  \to &
%  \bmat{cc|cc}{
%  I & O     &  O & O   \\
%  O & M_1   &  O & M_2 \\
%  O & M_3   &  O & M_4
%  },\\
%  &~~\text{symmetric $B+CA^T$ eliminated by phase and CZ gates,}\\
  = &
  \bmat{ccc|ccc}{
  I & O      & O          &    O & O      & O      \\
  O & M_{11} & M_{12}     &    O & M_{21} & M_{22} \\
  O & M_{31} & M_{32}     &    O & M_{41} & M_{42}
  }~~\text{by Lemma~\ref{IB}, and}\\
  &~~\text{recall that $M_1 = [M_{11}~M_{12}]$ with non-singular $M_{11}$, and by \eq{ccs0}:}
}
\beql{ccs1}{
  \bcase{
  M_{11} M_{21}^T + M_{12} M_{22}^T + M_{21} M_{11}^T + M_{22} M_{12}^T= O \\
  M_{31} M_{41}^T + M_{32} M_{42}^T + M_{41} M_{31}^T + M_{42} M_{32}^T= O \\
  M_{11} M_{41}^T + M_{12} M_{42}^T + M_{21} M_{31}^T + M_{22} M_{32}^T= I;
  }
}
\beqs{
  \to &
  \bmat{ccc|ccc}{
  I & O                 & O          &    O & O              & O      \\
  O & I                 & M_{12}     &    O & M_{21}M_{11}^T & M_{22} \\
  O & M_{31}M_{11}^{-1} & M_{32}     &    O & M_{41}M_{11}^T & M_{42}
  }\text{by Rule 1);}\\
%  &~~\text{where $M_{11}$ becomes $M_{11}M_{11}^{-1} = I$ by CNOT gates, and when}\\
%  &~~\text{the LHS times $M_{11}^{-1} = (P_1 P_2 \cdots)$ with  elementary column}\\
%  &~~\text{operations $P_i$, the RHS times $(P_1^T P_2^T \cdots) = (\cdots P_2 P_1)^T = M_{11}^T$}\\
%  &~~\text{as $P_i = P_i^{-1}$,}\\
  \\
  \to &
  \bmat{ccc|ccc}{
  I & O                 & O       &    O & O   & O      \\
  O & I                 & O       &    O & K_2 & M_{22} \\
  O & M_{31}M_{11}^{-1} & K_3     &    O & K_4 & M_{42}
  }\text{by Theorem~\ref{IA_IC},}\\
  &~~\text{where $M_{12}$ is eliminated by CNOT gates,}\\
  &~~\text{$K_2 = M_{21}M_{11}^T+M_{22}M_{12}^T$ is symmetric by \eq{ccs1},}\\
  &~~\text{$K_3 = M_{32}+M_{31}M_{11}^{-1}M_{12}$, and}\\
  &~~\text{$K_4 = M_{41}M_{11}^T+M_{42}M_{12}^T$;}\\
  \\
  \to &
  \bmat{ccc|ccc}{
  I & O                 & O       &    O & O    & O   \\
  O & I                 & O       &    O & K_2  & O   \\
  O & M_{31}M_{11}^{-1} & K_3     &    O & L_2  & L_4
  }\text{by Theorem~\ref{IA_IC},}\\
  &~~\text{where $M_{22}$ is eliminated by CZ gates, $L_2 = K_4+K_3 M_{22}^T$,}\\
  &~~\text{and $L_4 = M_{42}+M_{31}M_{11}^{-1}M_{22}$;}\\
  \\
  \to &
  \bmat{ccc|ccc}{
  I & O                 & O       &    O & O & O   \\
  O & I                 & O       &    O & O & O   \\
  O & M_{31}M_{11}^{-1} & K_3     &    O & I & L_4
  }\text{by Lemma~\ref{IB},}\\
  &~~\text{where the symmetric $K_2$ is eliminated by phase and CZ gates}\\
  &~~\text{and $L_2$ becomes $L_2 +  M_{31}M_{11}^{-1}K_2 = I$ by substituting}\\
  &~~\text{$K_2=K_2^T = M_{11}M_{21}^T+M_{12}M_{22}^T$,}\\
  \\
  \to &
  \bmat{ccc|ccc}{
  I & O & O       &    O & O                 & O   \\
  O & O & O       &    O & I                 & O   \\
  O & I & K_3     &    O & M_{31}M_{11}^{-1} & L_4
  }\text{by Rule 3),}\\
  &~~\text{where $\bsmtx{I\\ M_{31}M_{11}^{-1}}$ and $\bsmtx{O\\ I}$ are swapped by Hadamard gates;}\\
  \\
  \to &
  \bmat{ccc|ccc}{
  I & O & O     &    O & O & O   \\
  O & O & O     &    O & I & O   \\
  O & I & O     &    O & W & L_4
  }\text{by Theorem~\ref{IA_IC},}\\
  &\text{where $K_3$ is eliminated by CNOT gates and }\\
  &\text{$W =  M_{31}M_{11}^{-1} + L_4 K_3^T$ is symmetric as follows:}\\[-22pt]
}
\beqs{
\quad
&W = M_{31}M_{11}^{-1} (I+M_{22}M_{32}^T) +M_{42}M_{32}^T \\
&~~~~ +M_{42}M_{12}^T(M_{31}M_{11}^{-1})^T +M_{31}M_{11}^{-1}M_{22}M_{12}^T(M_{31}M_{11}^{-1})^T,\\
&\text{and by \eq{ccs1}, $(I+M_{22}M_{32}^T) = M_{11} M_{41}^T + M_{12} M_{42}^T + M_{21} M_{31}^T$,}\\
&\text{and  only symmetric terms are left in $W$;}
}
\beqs{
  \to &
  \bmat{ccc|ccc}{
  I & O & O    &    O & O & O \\
  O & O & O    &    O & I & O \\
  O & I & O    &    O & O & O
  }\text{by Theorem~\ref{IA_IC} and Lemma~\ref{IB},} ~~~~~\\ %squeenze the matrix to left
  &~~\text{where $L_4$ and the symmetric matrix $W$ are eliminated by CZ gates;}\\
  \\
  \to &
  \bmat{ccc|ccc}{
  O & O & O    &    I & O & O \\
  O & I & O    &    O & O & O \\
  O & O & O    &    O & I & O
  }_{(s+2c)\times 2n} \text{by Rule 3),}\\
  &~~\text{where Hadamard gates are applied to qubits $1,2,\dots, s+c$.}
}
\normalsize
We have shown that, to reduce \eq{H'} to \eq{H'_raw}, the required number of gates is proportional to $(n-k+c) \times 2n$.
By a deduction similar to the proof of Theorem~\ref{O(H)} (cf. the mentioned trick prior to this theorem),
the overall complexity is $O(n(n-k+c)/\log n)$.
Notice that, when $c$ is large, a process like \eq{B_0} is required to reduce $M_{11}$ to $I$ efficiently.
\end{proof}

\section{Hardness of decoding EAQSCs} \label{sec:DecComp}
%We first sketch that a major step (finding a proper error) for decoding an $[[n,k;c]]$ EA stabilizer code can be considered as a classical computational problem.
%Recall that the mapping $\varphi$ in Section \ref{sec:StdForm} makes us be able to work in binary vector space.
We first describe the decoding procedure of an $[[n,k;c]]$ EAQSC
with a standard-form check matrix $H$ in \eq{hH}. 
Suppose that a Pauli error $E   \in \sG_{n+c}$
occurs during the transmission of a codeword.
The error syndrome vector is defined by
$y = \varphi(E)\La H^T \in \ZZ_2^{(s+2c)}$ %after projective measurements by stabilizers 
\cite{KL13}. 
%provided that the error is $E = E'\otimes I^{\otimes c} \in \sG_{n+c}$ for some error $E' \in \sG_n$ caused by the channel. 
Since Bob's original qubits are error-free by assumption, 
$E = E'\otimes I^{\otimes c} \in \sG_{n+c}$ for some error $E' \in \sG_n$. 
Thus we have $y = \varphi(E')\La H'^T$ where $H'$ is the corresponding simplified check matrix as in \eq{H'}.
Given $y$, the receiver (Bob) has to find a proper error vector $e\in\ZZ_2^{2n}$ such that $e\La H'^T = y$.
Then  a correction operator $\hE\in \varphi^{-1}(e)\otimes I^{\otimes c}$ is applied.

Two types of decoders are considered regarding to the degeneracy of EAQSCs.
The quantum maximum likelihood decoder finds a  minimum-weight error with syndrome $y$.
If degeneracy is considered, the (optimal) degenerate quantum maximum likelihood decoder finds the most probable coset of degenerate errors with syndrome $y$~\cite{HG10,KL13}.
For example, consider the independent $X$--$Z$ channel model, where a qubit suffers an $X$ error  with probability $p$ and independently it suffers a $Z$ error with probability $p$ for $p\in [0, 0.5)$.
 The \emph{coset probability} of an $e\in\ZZ_2^{2n}$ with respect to a check matrix $H'$ 
 is defined as 
\beqsl{P(eH)}{
	P(e+\Row(H')) &= \sum_{u\in e+\Row(H')} P(u) \\
	&= \sum_{u\in e+\Row(H')} p^{\wt(u)}(1-p)^{2n-\wt(u)},
}
where $\Row(H')\subset \ZZ_2^{2n}$ is the row space of $H'$, $\wt(u)$ is the Hamming weight of $u\in\ZZ_2^{2n}$, and $P(u) = p^{\wt(u)}(1-p)^{2n-\wt(u)}$ is the probability that an error in $\varphi^{-1}(u)$ occurs.
%Normally, a lower-weight error has a higher probability to occur, but with \emph{degeneracy}, an error with a higher \emph{coset probability} becomes more worth to correct \cite{HG10}, \cite{KL13}.

%Considering the entanglement assistance, we state two decoding problems and use previous results to show their hardnesses.

Previously the quantum maximal-likelihood decoding (QMLD) of general QSCs is known to be NP-hard
for the independent $X$--$Z$ channel~\cite{HG10} and depolarizing channel~\cite{KL13}. 
Moreover, the   degenerate quantum maximum likelihood decoding (DQMLD) of QSCs is shown to be \#P-complete~\cite{IP15}.
%by  a weight enumerator counting problem (formed as a decision problem) in polynomial time 
We would like to generalize these results to EAQSCs with respect to the independent $X$--$Z$ channel.
%=========================
\\[8pt]
{\bf \textsc{EAQSC Maximum Likelihood Decoding (EMLD)} }\\[2pt]
\textsc{Input:}  A full row-rank $(s+2c)\times 2n$ binary matrix $H'$ satisfying the requirements in \eq{ccs} and \eq{H'}, and a binary vector $y\in\ZZ_2^{s+2c}$.\\[1pt]
\textsc{Output:} A binary vector $e\in\ZZ_2^{2n}$ satisfying $e\La H'^T = y$
                 and minimizing $\wt(e)$.
\\[-8pt]
%=========================
\\[8pt]
{\bf \textsc{Degenerate EAQSC Maximum Likelihood Decoding (DEMLD)} }\\[2pt]
\textsc{Input:}  A full row-rank $(s+2c)\times 2n$ binary matrix $H'$ satisfying the requirements in \eq{ccs} and \eq{H'}, a binary vector $y\in\ZZ_2^{s+2c}$, and a real number $ p\in [0, 0.5)$.\\[1pt]
\textsc{Output:} A binary vector $e\in\ZZ_2^{2n}$ satisfying $e\La H'^T = y$
                 and maximizing the coset probability $P(e+\Row(H'))$ in \eq{P(eH)}.
\\[-2pt]
%=========================
\bc \label{NP} %%%%%%%%%%%%%%%%%%%%%%%%%%%%%%%%
EMLD is NP-complete.
\ec %%%%%%%%%%%%%%%%%%%%%%%%%%%%%%%%%%%%%%%%%%
\begin{proof}	
A general EAQSC decoder  supports the case $c=0$. Thus   the QMLD in \cite{HG10} trivially reduces to EMLD in polynomial time. By the main theorem of \cite{HG10}, EMLD is NP-complete.
\end{proof}
%%%%%%%%%%%%%%%%%%%%%%%%%%%%%%%%%%%%%%%%%%%
\bc \label{sharpP} %%%%%%%%%%%%%%%%%%%%%%%%%%%%%%%%
DEMLD is \#P-complete.
\ec %%%%%%%%%%%%%%%%%%%%%%%%%%%%%%%%%%%%%%%%%%
\begin{proof}	
Consider $c=0$ and $k=1$. Then  DQMLD defined in \cite{IP15} reduces to  DEMLD in polynomial time. By Theorem~2 of \cite{IP15}, DEMLD is \#P-complete.
\end{proof}
%%%%%%%%%%%%%%%%%%%%%%%%%%%%%%%%%%%%%%%%%%%
It is straightforward to extend the results to the case of depolarizing channel by redefining the $P(u)=P(x|z)$ in \eq{P(eH)} and in DEMLD as 
$P(u) = (p/3)^{\gw(u)}(1-p)^{n-\gw(u)}$, where $0\le p< 3/4$ and $\gw(x|z) \teq \wt(x\vee z)$ is the \emph{generalized weight}, where $x\vee z$  is the bitwise OR of $x$ and $z$. By Theorem~5 of \cite{KL13} and the conclusion remark in~\cite{IP15}, we have the following remark:
\br \label{DepCh} %%%%%%%%%%%%%%%%%%%%%%%%%%%%%%%%
Corollaries \ref{NP} and \ref{sharpP} also hold   for the depolarizing channel with respect to generalized weight.
\er %%%%%%%%%%%%%%%%%%%%%%%%%%%%%%%%%%%%%%%%%%
%These results could be used to generalize the quantum McEliece cryptosystem \cite{Fuj12} as an EAQSC McEliece cryptosystem.

\section{Conclusion} \label{sec:Conclu}
We have improved the encoding complexity to $O(n(n-k+c)/\log n)$ for an $[[n,k;c]]$ EAQSC. This also suggests that entanglement benefits an EAQSC at the cost of additional encoding complexity.
% but gains the benefit of a larger minimum distance.
On the other hand, we showed that decoding general EAQSC is \#P-complete  if degeneracy is considered
and NP-complete if not. 
%This potentially leads to the application of a secure McEliece cryptosystem from EA stabilizer codes.

CYL was  financially supported from the Young Scholar Fellowship Program by Ministry of Science and Technology (MOST) in Taiwan, under
Grant MOST107-2636-E-009-005.

\end{document}